\documentclass[conference]{IEEEtran}
\IEEEoverridecommandlockouts
\usepackage{cite}
\usepackage{amsmath,amssymb,amsfonts}
\usepackage{algorithm}
\usepackage{algpseudocode}
\usepackage{graphicx}
\usepackage{textcomp}
\usepackage{xcolor}
\def\BibTeX{{\rm B\kern-.05em{\sc i\kern-.025em b}\kern-.08em
    T\kern-.1667em\lower.7ex\hbox{E}\kern-.125emX}}
 \usepackage{subcaption}
 \usepackage{url}
\newcommand{\bm}{\mathbf}
\newcommand{\be}{\begin{equation}}
\newcommand{\ee}{\end{equation}}
\newcommand{\bea}{\begin{eqnarray}}
\newcommand{\eea}{\end{eqnarray}}
\newcommand{\bb}{{\bm b}}
\newcommand{\bc}{{\bm c}}
\newcommand{\x}{{\bm x}}

\newcommand{\bF}{{\bf F}}

\newcommand{\bH}{{\bf H}}

\newcommand{\blambda}{\mbox{\boldmath$\lambda$}}

\newcommand{\bnu}{\mbox{\boldmath$\nu$}}

\newcommand{\bmu}{\mbox{\boldmath$\mu$}}
\newcommand{\balpha}{\mbox{\boldmath$\alpha$}}
\newcommand{\bDelta}{\mbox{\boldmath$\Delta$}}

\begin{document}
\title{Truncated Turbo Equalizer with SIC for OTFS}
\author{\IEEEauthorblockN{Sanoopkumar P. S., Stephen McWade, Arman Farhang}
\IEEEauthorblockA{Department of Electronic \& Electrical Engineering, Trinity College Dublin, Ireland \\
\{pungayis, smcwade, arman.farhang\}@tcd.ie.\vspace{-5 mm}}}
\maketitle
\begin{abstract}

Orthogonal time frequency space (OTFS) is a promising candidate waveform for the next generation wireless communication systems. OTFS places data in the delay-Doppler (DD) domain, which simplifies channel estimation in high-mobility scenarios. However, due to the 2-D convolution effect of the time-varying channel in the DD domain, equalization is still a challenge for OTFS. Existing equalizers for OTFS are either highly complex or they do not consider intercarrier interference present in high-mobility scenarios. Hence, in this paper, we propose a novel two-stage detection technique for coded OTFS systems. Our proposed detector brings orders of magnitude computational complexity reduction compared to existing methods. At the first stage, it truncates the channel by considering only the \textit{significant} coefficients along the Doppler dimension and performs turbo equalization. To reduce the computational load of the turbo equalizer, our proposed method deploys the modified LSQR (mLSQR) algorithm. At the second stage, with only two successive interference cancellation (SIC) iterations, our proposed detector removes the residual interference caused by channel truncation. To evaluate the performance of our proposed truncated turbo equalizer with SIC (TTE-SIC), we set the minimum mean squared error (MMSE) equalizer without channel truncation as a benchmark. Our simulation results show that the proposed TTE-SIC technique achieves about the same bit error rate (BER) performance as the benchmark. 
\end{abstract}

\section{Introduction}
In recent years, orthogonal time frequency space modulation (OTFS) has emerged as a highly promising candidate waveform for the next generation wireless communication systems \cite{hadani_2017}. OTFS is an attractive candidate due to its high resilience to the Doppler spread in high-mobility environments, where the wireless channel is time-varying.  In contrast to orthogonal frequency division multiplexing (OFDM), which multiplexes the modulated data symbols in the time-frequency (TF) domain,  OTFS transmits data symbols in delay-Doppler (DD) domain \cite{Farhang_letter}.
The channel in the DD domain varies much more slowly than in the TF domain, which allows for channel estimation to be performed with a small pilot overhead, even in high-mobility scenarios \cite{ps2022practical}.  Despite this, channel equalization and data detection for OTFS are challenging tasks due to the 2D convolution effect of the time-varying channel \cite{zhang2022survey}.

There exists a large body of literature on equalization and detection for OTFS \cite{Surabhi_2020,Raviteja_2018,Li_2021,Qu_lsmr,Zemen_2018,Long_2019,Das_2020}. However, only a small number of these works consider a coded system \cite{Zemen_2018,Long_2019,Das_2020}, which is the topic of interest to this paper. 
With regards to coded OTFS, the authors of \cite{Zemen_2018} developed an iterative parallel interference cancellation (PIC) detection scheme applicable to orthogonal precoded TF modulations.  An iterative linear minimum mean squared error with PIC (LMMSE-PIC) equalizer was developed in \cite{Long_2019}. This method makes use of component-wise conditionally unbiased LMMSE estimator (CWCU-LMMSE) and first order Neumann series approximation. However, in \cite{Zemen_2018} and \cite{Long_2019}, the channel is assumed to be locally time invariant across a block of samples. In high-mobility scenarios, where the channel is highly time-varying, due to the presence of inter-carrier interference (ICI), this assumption becomes invalid. Therefore, these methods suffer from performance penalty or MMSE equalization becomes highly complex in tackling the ICI issue.
In \cite{Das_2020}, the authors investigated the performance of low-density parity check (LDPC) coded OTFS. However, the technique proposed in \cite{Das_2020} relies on hard decision output from the LDPC decoder, and is therefore prone to error propagation. The authors of \cite{Li_2023_turbospars} proposed a turbo receiver which uses a sparsified correlation matrix to implement MMSE equalization with reduced complexity. However, this method uses the computationally complex factorized sparse approximate inverse (FSPAI) algorithm which limits its use in practical scenarios.  
\begin{figure*}[t]
\vspace{0 cm}
\centering
\includegraphics[scale=0.65]{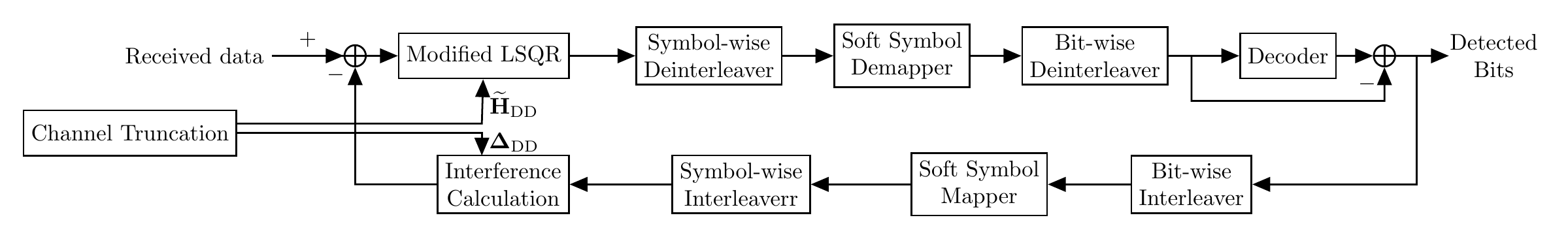}
\vspace{-0.3 cm}
\caption{Block diagram of the proposed TTE-SIC receiver.}\label{turbo_sic_blkdia}
\vspace{-0.5 cm}
\end{figure*}

Hence, the existing methods in the literature do not provide a low complexity practical data detection method in high-mobility scenarios. To address this shortcoming, we propose a novel low-complexity truncated turbo equalizer with successive interference cancellation (TTE-SIC) for coded OTFS. Our proposed method has two stages; in the first stage, we reduce the equalization complexity by truncating the channel. For channel truncation, we introduce a criterion based on the maximum Doppler shift. To further reduce complexity, we utilize the low-complexity modified LSQR (mLSQR) algorithm from \cite{mcwade2023lowcomplexity}. With this approach, we obtain the soft estimate of the transmitted symbols and the post-equalization signal to interference noise ratio (SINR) information. Using the output from mLSQR, we derive expressions for the log-likelihood ratios (LLRs) for soft-output detection. In the second stage, we compensate for the performance loss due to channel truncation with a low-complexity SIC procedure using the channel coefficients remaining after truncation. We evaluate our proposed method via simulations and show that after only two SIC iterations, it achieves similar bit error rate (BER) performance as our benchmark, i.e., MMSE equalization without channel truncation. We analyze and compare the computational complexity of our proposed technique with the existing literature. We calculate expressions for computational load in terms of the number of complex multiplications (CMs). We show that our proposed technique has around $10^4$ and $10$ times lower complexity than our benchmark and the simplest method in the literature, respectively.

The rest of this paper is organized as follows. In Section \ref{sec:intro}, we introduce the system model under consideration. The proposed TTE-SIC and its computational complexity analysis are provided in Section \ref{sec:proposal}. The simulation results are presented in Section \ref{sec:simultion} and the paper is concluded in Section \ref{sec:conclusion}. 

\textit{Notations}: Superscripts ${(\cdot)^{\rm{T}}}$ and ${(\cdot)^{\rm{H}}}$ denote transpose and Hermitian, respectively. Bold lower-case characters are used to denote vectors and bold upper-case characters are used to denote matrices. $x[n]$ denotes the $n$-th element of the vector $\mathbf{x}$. The function $\rm{vec}(\mathbf{X})$ vectorizes the matrix $\mathbf{X}$ by stacking its columns to form a single column vector, and $\otimes$ represents the Kronecker product. The $p\times{p}$ identity matrix and $p \times q$ all-zero matrix are  denoted by $\mathbf{I}_p$ and $\mathbf{0}_{p\times{q}}$, respectively.

\section{System Model}\label{sec:intro}
We consider a coded OTFS system with $M$ delay bins, $N$ Doppler bins, and a $\frac{k}{r}$ rate forward error correction (FEC) code.  At the transmitter, the FEC encoder converts the information bits, $\bb\in \{0,1\}^{MNk}$ to $\bc\in \{0,1\}^{MNr}$ coded bits.  After bit-wise interleaving, the coded bits are mapped onto  $Q=2^r-$QAM modulation constellation to form the DD domain transmit data symbols  $\x\in\mathbb{C}^{MN\times 1}$. After symbol-wise interleaving, the DD domain data symbols are rearranged as the $M \times N$ matrix $\bm{X}$. 
In this paper, we consider the full cyclic prefix (CP) OTFS system in which a CP is appended at the beginning of each block of $M$ samples in the delay-time (DT) domain. The DT domain  transmit signal is formed by taking $N$-point IDFT across the Doppler dimension, i.e., the rows of $\bm{X}$.  Therefore, the OTFS transmit signal is given by $\bm{S} = {\bm{A}_{\rm{cp}}}\bm{X}\bm{F}_{N}^{\rm{H}},$ where $\bm{F}_N$ is the $N$-point unitary discrete Fourier transform (DFT) matrix with $(l,k)$ elements $\frac{1}{\sqrt{N}}e^{-j\frac{2\pi}{N}lk}$ for $l,k=0,\ldots,N-1$. ${\bm{A}_{\rm{cp}}} = \left[\bm{J}_{\rm{cp}}, \bm{I}_{M}\right]$ is the CP addition matrix where $\bm{J}_{\rm{cp}}$ is composed of the last $M_{\mathrm{cp}}$ rows of $\mathbf{I}_{M}$.
After parallel to serial conversion, the time-domain signal is represented in vectorized form as 
\be
    \bm{s} = {\rm vec}(\bm{S}) = (\bm{F}_{N}^{\rm{H}} \otimes \bm{A}_{\rm{cp}})\bm{x}, \label{eq2}
\ee
where $\bm{x} = \rm{vec}(\bm{X})$. 
The transmit signal then undergoes analog to digital conversion and is transmitted through the LTV channel. The continuous time received signal is thus represented by $r(t) = \int \int h(\tau,\nu)s(t-\tau)e^{j2\pi\nu(t-\tau)}d\tau d\nu + \eta(t)$,
%
%
where $h(\tau,\nu) = \sum_{p=0}^{P-1}h_{p}\delta(\tau - \tau_{p})\delta(\nu - \nu_{p}),$ is the DD domain channel impulse response (CIR), which consists of $P$ channel paths, and $\eta(t)$ is the complex additive white Guassian noise (AWGN) with variance $\sigma^2$. The parameters $h_{p}$, $\tau_{p}$ and $\nu_{p}$ represent the channel gain, delay and Doppler shift associated with path $p$ of the channel,  respectively.

The received signal is then sampled with sampling period $T_{\rm{s}}$ to obtain the discrete-time received signal, given by
\be
    \bm{r} = \bm{H}_{\rm{DT}}\bm{s} + \boldsymbol{\eta}_{\rm{DT}},
\ee
where $\boldsymbol{\eta}_{\rm{DT}}$ is the $MN \times 1$ AWGN vector and $\bm{H}_{\rm{DT}}$ is the $MN \times MN$ DT domain channel matrix. 
The received signal is then demodulated and converted back to the DD domain by performing an $N$-point DFT operation across the time-domain samples. Thus, the received signal is given by $\mathbf{y} = (\mathbf{F}_{N}\otimes\mathbf{R}_{\mathrm{cp}})\mathbf{r}.$ This can be alternatively expressed as
\be\label{eqn:y_dd}
    \bm{y}_{\rm DD} = \bm{H}_{\rm{DD}}\bm{x} + \boldsymbol{\eta}_{\rm{DD}},
\ee
where $\bm{H}_{\rm{DD}} = (\bm{F}_N \otimes \bm{R}_{\rm{cp}})\bm{H}_{\rm{DT}}(\bm{F}_N^{\rm{H}} \otimes \bm{A}_{\rm{cp}})$ is the effective DD domain channel matrix, $\bm{R}_{\rm{cp}} = \left[\bm{0}_{M\times M_{\rm{cp}}}, \bm{I}_{M}\right]$ is the CP removal matrix and $\boldsymbol{\eta}_{\rm{DD}}=(\bm{F}_{N}\otimes\bm{R}_{\rm{cp}})\boldsymbol{\eta}_{\rm{DT}}$ is the DD domain noise vector. 

\section{Proposed Joint Equalization and Decoding}\label{sec:proposal}
In this section, we present our proposed equalization and detection technique, as shown in Fig. \ref{turbo_sic_blkdia}.  Our proposed technique has two-stages. In the first stage, we reduce the equalization complexity by truncating the DD domain channel matrix along the Doppler dimension to only include the \textit{significant} Doppler coefficients. In the second stage, we deploy SIC to remove the residual interference caused by the channel truncation. Furthermore, in the first stage, we use the low-complexity mLSQR algorithm from \cite{mcwade2023lowcomplexity} to equalize the channel, obtain the soft estimate of the transmitted symbols and find the post-equalization SINR information. We use the output of the mLSQR algorithm to derive expressions for the extrinsic log-likelihood ratios for soft-output detection. In the following subsections, we provide a detailed explanation of the stages of our proposed technique, beginning with the channel truncation.

\subsection{Channel truncation}\label{sec_truncation}
At the OTFS receiver, linear equalization techniques such as least square (LS) and MMSE can be performed in either the DT, TF or DD domains. However, these methods are not viable solutions for practical implementation due to the computational complexity of computing the inverse of the $MN \times MN$ channel matrix. To reduce the equalization complexity, we truncate the channel along the Doppler dimension. This truncation naturally results in performance loss for the linear equalizer which we compensate for with interference cancellation later in the procedure using the soft output from the decoder. 

\begin{figure}[t]
\centering
\begin{subfigure}{0.5\columnwidth}
	\centering
	\includegraphics[width=0.5\columnwidth]{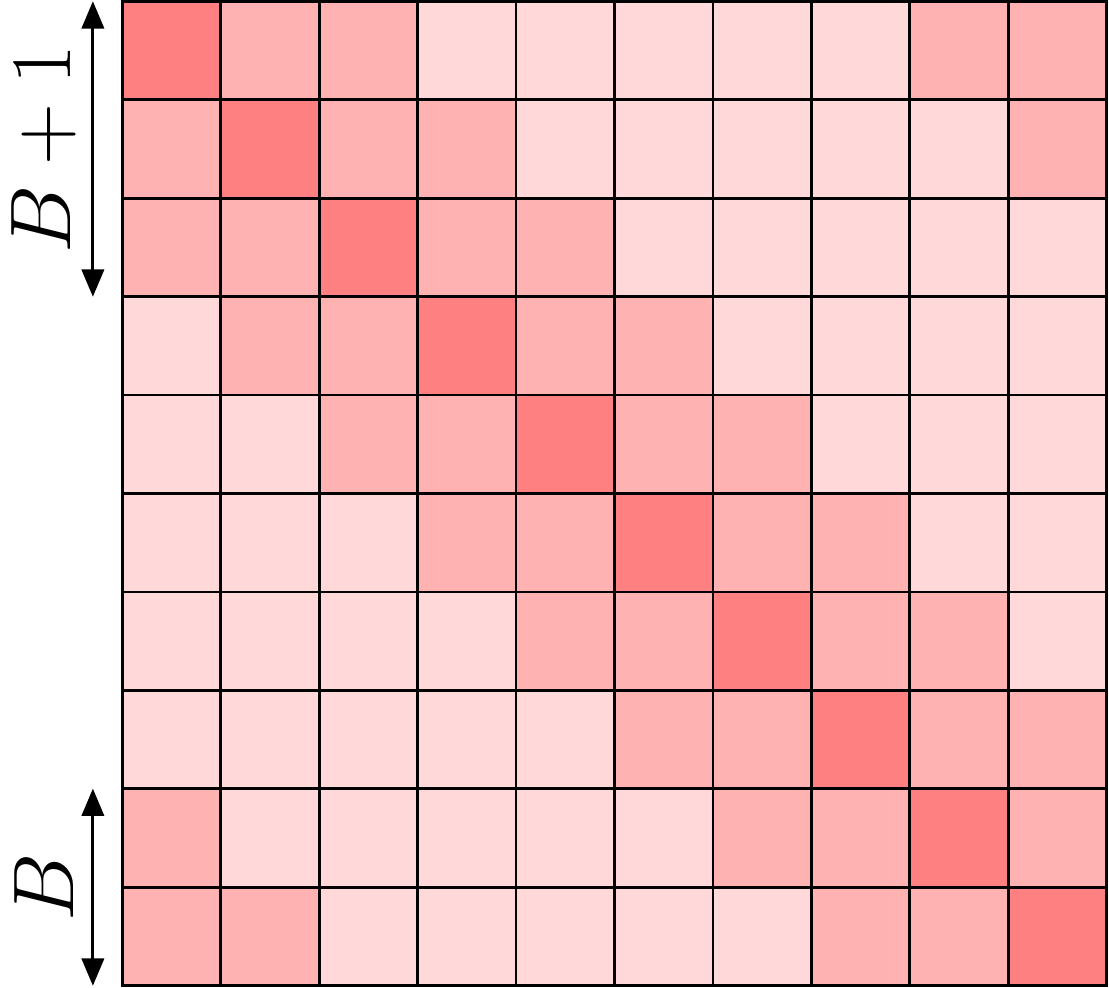}
	\caption{${\bm H}_{\bm {DD}}$}
	\label{fig:channel_dd}
\end{subfigure}
\hspace{-1 cm}
\begin{subfigure}{0.5\columnwidth}
	\centering
	\includegraphics[width=0.5\columnwidth]{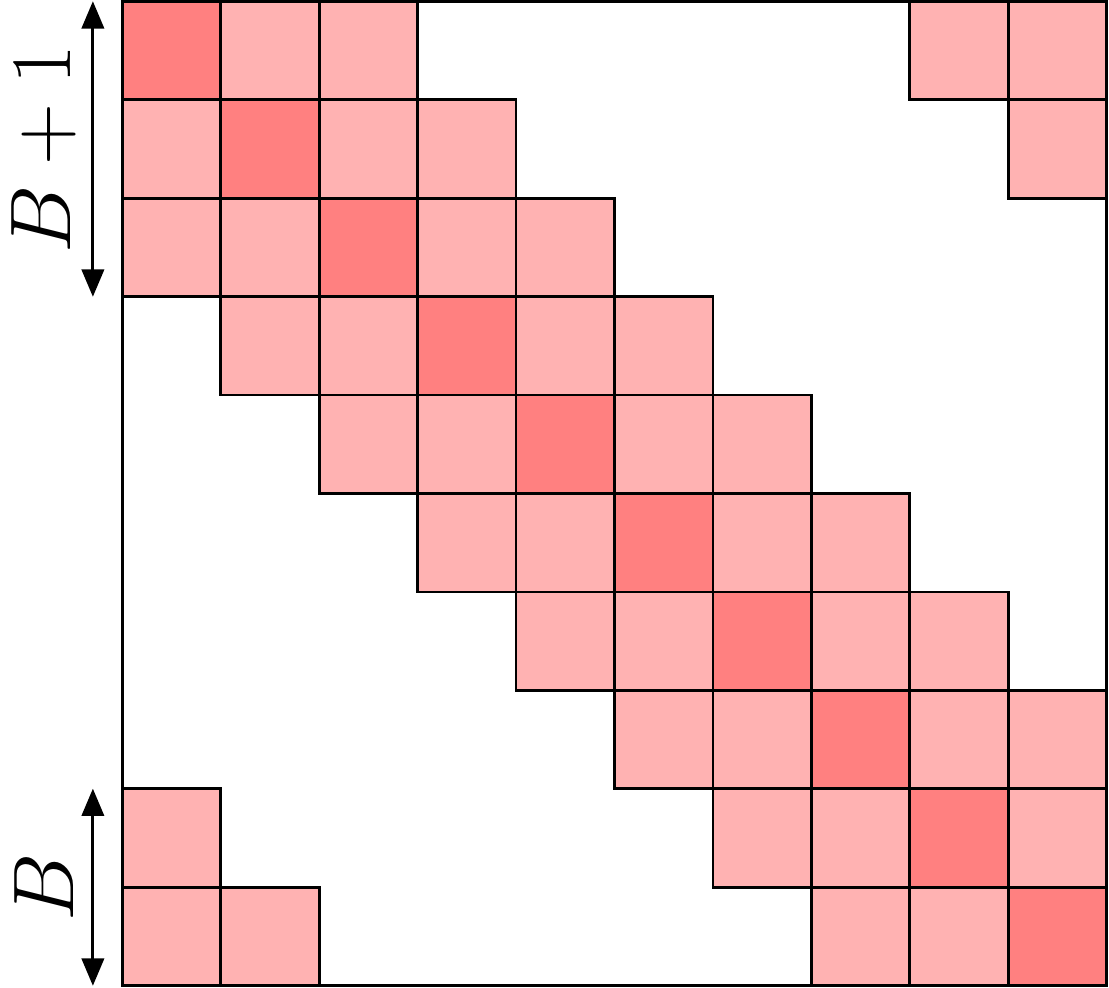}
	\caption{ $\widetilde{\bH}_{\rm DD}$}
	\label{fig:channel_sub_dd}
\end{subfigure}
\vspace{-0.2cm}
\caption{Blockbanded structure ${\bm H}_{\bm {DD}}$ and $\widetilde{\bH}_{\rm DD}$.}
\label{fig:channel_matrix}
\vspace{-0.5 cm}
\end{figure}

The DD domain channel matrix $\bH_{\rm DD}$, as seen in Fig.~\ref{fig:channel_dd}, is a block circulant matrix with $M\times M$ subblocks. In Fig.~\ref{fig:channel_matrix} each small square represents a $M \times M$ submatrix.  In low-mobility scenarios, $\bH_{\rm DD}$ can be assumed to be a block circulant matrix with circulant blocks (BCCB). Furthermore,  the structure of $\bH_{\rm DD}$ can be viewed as a block banded matrix where the diagonal subblocks have the most significant magnitude and the magnitude of elements in off-diagonal subblocks reduces as we move away from the main diagonal submatrices. 

We define the \textit{significant} subblocks of $\bH_{\rm DD}$ as the subblocks at each Doppler index which are contained within the truncation bandwidth, $B=\lceil f_{\rm Dmax}MNT_{\rm s}\rceil$, where $f_{\rm Dmax}$ is the maximum Doppler shift and $T_{\rm s}$ is the sampling period. Thus, with the channel truncation, we consider only $2B+1$ \textit{significant} Doppler coefficients for channel equalization.  The \textit{insignificant} subblocks of $\bH_{\rm DD}$  are defined as those not contained within the truncation bandwidth. Hence, we decompose $\bH_{\rm DD}$ as $\bH_{\rm DD}=\widetilde{\bH}_{\rm DD}+\bDelta_{\rm DD}$,
where $\widetilde{\bH}_{\rm DD}$  and $\bDelta_{\rm DD}$ are the DD domain channel matrices with \textit{significant}  and \textit{insignificant} Doppler indices, respectively. We now rewrite (\ref{eqn:y_dd}) as
\be\label{eqn:y2}
\bm{y}_{\rm DD}=\widetilde{\bH}_{\rm DD}\x+\bDelta_{\rm DD}\x+\boldsymbol{\eta}_{\rm{DD}}.
\ee
In the proposed receiver, we use SIC to remove the interference caused by $\bDelta_{\rm DD}$ based on the soft output from the decoder. Based on the assumption that the soft output becomes more accurate after every iteration, we can approximate (\ref{eqn:y2}) as,
\be\label{eqn:y_app}
\bm{y}_{\rm DD}\approx \widetilde{\bH}_{\rm DD}\x+\boldsymbol{\eta}_{\rm{DD}}.
\ee
Based on (\ref{eqn:y_app}), in our proposed method, we use the mLSQR algorithm to equalize the truncated channel and to obtain the soft output based on the MMSE criteria \cite{Tuchler_2002}. The computational complexity of mLSQR is directly related to the number of non-zero elements in the channel matrix. By utilising the truncated channel matrix, we reduce the computational complexity of equalization further by using the LSQR algorithm.

\subsection{Modified LSQR Algorithm}
The proposed method uses the mLSQR algorithm to equalize the channel, which is listed in Algorithm 1. LSQR is a well-known iterative algorithm for solving problems of the form $\bm{y}_{\rm DD}=\widetilde{\bH}_{\rm DD}\x+\boldsymbol{\eta}_{\rm{DD}}$, \cite{Hrycak_LSQR_GMRES}. 
At each iteration, $k$, LSQR uses Golub-Kahan bidiagonalization and QR decomposition to obtain an estimate of the transmitted symbol $\mathbf{x}_k$ \cite{Hrycak_LSQR_GMRES}. A simple recursive method for updating this estimate within each iteration was proposed in \cite{LSQR}. The iterative process continues until either the norm of the residual reaches a pre-determined tolerance, $\epsilon$, or the maximum number of iterations $I_{\rm LSQR}$ is reached. LSQR can also be regularized by including the noise variance $\sigma^2$ as a damping parameter to improve the semi-convergence property of the algorithm. After several iterations, LSQR provides a performance similar to MMSE but with a substantially lower complexity \cite{Hrycak_LSQR_GMRES}.
\begin{algorithm}[t]
\caption{Modified LSQR Algorithm}\label{LSQR}
\begin{algorithmic}[1]
\State \textbf{Input}: $\widetilde{\bH}_{\rm DD}$, $\mathbf{y}_{\rm DD}$ and $\sigma^2$
\State \textbf{Initialize}: $\Bar{\mathbf{y}}=[
  \mathbf{y}_{\rm DD}^{\rm{T}},
  \mathbf{0}^{\rm{T}}]^{\rm{T}}$, $\mathbf{A}_{\rm DD} = [\widetilde{\bH}_{\rm{DD}}^{\rm{T}}, \sigma\mathbf{I}_{MN}^{\rm{T}}]^{\rm{T}}$, $\beta_0 = \|\Bar{\mathbf{y}}\|$, $\mathbf{u}_0 = \Bar{\mathbf{y}}/\beta_0$, $a_0=\|\mathbf{A}_{\rm DD}^\mathrm{H}\mathbf{u}_0\|$, $\mathbf{v}_0=\mathbf{A}_{\rm DD}^\mathrm{H}\mathbf{u}_0/a_0$, $\mathbf{w}_0 = \mathbf{v}_0$, $\Bar{\phi}_0 = \beta_0$, $\Bar{\rho}_0 = a_0$, $\mathbf{x}_0 = \mathbf{0}_{MN \times 1}$, $\mathbf{W}_{1} = \frac{\zeta_1}{\Bar{\rho}_{0}\Bar{\phi}_{0}}\mathbf{I}_{\mathrm{MN}}$, $\mathbf{W}_0 = \mathbf{0}_{MN \times MN}$, $\zeta_0=1$ and $\Bar{\phi}_k=\Bar{\rho}_k=1$ for $k<0$
\For{$k=1:I_{\rm LSQR}$}
  \State $\beta_k = \|\mathbf{A}_{\rm DD}\mathbf{v}_{k-1}-a_{k-1}\mathbf{u}_{k-1}\|$
  \State $\mathbf{u}_k = (\mathbf{A}_{\rm DD}\mathbf{v}_{k-1}-a_{k-1}\mathbf{u}_{k-1})/\beta_k$
  \State $a_k = \|\mathbf{A}_{\rm DD}^\mathrm{H}\mathbf{u}_k-\beta_k\mathbf{v}_{k-1}\|$
  \State $\mathbf{v}_k = (\mathbf{A}_{\rm DD}^\mathrm{H}\mathbf{u}_k-\beta_k\mathbf{v}_{k-1})/a_k$ 
  \State $\rho_k = \|[\Bar{\rho}_{k-1}\ \ \beta_k]\|$, $z_k = \frac{\Bar{\rho}_{k-1}}{\rho_k}$, $s_k = \frac{\beta_k}{\rho_k}$, $\theta_k = s_ka_k$
  \State $\phi_k = z_k\Bar{\phi}_{k-1}$, $\zeta_k = \frac{\phi_k}{\rho_k}$, $\psi_k = \frac{\theta_k}{\rho_k}$
  \State $\Bar{\phi}_k = -s_k\Bar{\phi}_{k-1}$, $\Bar{\rho}_k = -z_k\Bar{\mathbf{y}}_k$
  \State $\mathbf{x}_k = \mathbf{x}_{k-1} + \zeta_k\mathbf{w}_{k-1}$
  \State $\mathbf{w}_k=\mathbf{v}_k- \psi_k\mathbf{w}_{k-1}$
  \State Compute $\mathbf{\Omega}_k$ using (\ref{eq10_apx})
  \State if $||\mathbf{y} -\widetilde{\bH}_{\rm DD}\mathbf{x}_k || \leq \epsilon$, break
\EndFor
\State Compute $\mu_{\hat{\x}}[n]$ and $\nu_{\hat{\x}}[n],\ $ $\forall n$ using (\ref{mean_apx}) and (\ref{var_apx})
\State \textbf{Output}: $\hat{\x} = \mathbf{x}_k$,  $\bmu_{\hat{\x}}$ and $\bnu_{\hat{\x}}$
\end{algorithmic}
\end{algorithm}
We note that the conventional LSQR algorithm does not provide the post-equalization SINR information which is required for computing the LLRs. Therefore, we use the modified LSQR algorithm from \cite{mcwade2023lowcomplexity}, to obtain this information. LSQR computes $\mathbf{x}_k$ at each iteration using a simple recursion. However, similar to the conjugate gradiant method in \cite{Yin_CG_mod}, $\mathbf{x}_k$ can also be computed using an LSQR equivalent equalization matrix which depends on the iteration index $k$. The LSQR equivalent equalization matrix at iteration $k$ is defined as $\mathbf{W}_k\widetilde{\bH}_{\rm DD}^{\mathrm{H}}$, and $\mathbf{x}_k$ can be written as
    $\mathbf{x}_k = \mathbf{W}_k\widetilde{\bH}_{\rm DD}^{\mathrm{H}}\mathbf{y}$. 
 From \cite{mcwade2023lowcomplexity}, $\mathbf{W}_k$ can be obtained recursively as
\begin{equation}
    \begin{split}
        &\mathbf{W}_k  =  \mathbf{W}_{k-1} + \frac{\psi_{k-2}^2\zeta_k\Bar{\rho}_{k-3}\Bar{\phi}_{k-3}}{\zeta_{k-2}\Bar{\rho}_{k-1}\Bar{\phi}_{k-1}}\left(\mathbf{W}_{k-2} -\mathbf{W}_{k-3}\right) +
        \\& \left( \frac{\zeta_k\Bar{\rho}_{k-2}\Bar{\phi}_{k-2}(1+\psi_{k-1}^2)}{\zeta_{k-1}\Bar{\rho}_{k-1}\Bar{\phi}_{k-1}}\mathbf{I}_{MN} - \frac{\zeta_k}{\Bar{\rho}_{k-1}\Bar{\phi}_{k-1}}\mathbf{A}_{\rm DD}^\mathrm{H}\mathbf{A}_{\rm DD} \right)
        \\& \times \left(\mathbf{W}_{k-1} -\mathbf{W}_{k-2}\right),
    \end{split}
        \label{eq10}
\end{equation}
where $\mathbf{A}_{\rm DD} = [\widetilde{\bH}_{\rm{DD}}^{\rm{T}}, \sigma\mathbf{I}_{MN}^{\rm{T}}]^{\rm{T}}$, $\mathbf{W}_{1} = \frac{\zeta_1}{\Bar{\rho}_{0}\Bar{\phi}_{0}}\mathbf{I}_{MN}$, $\mathbf{W}_k = \mathbf{0}_{MN \times MN}$ for $k \le 0$, $\zeta_0=1$ and $\Bar{\phi}_k=\Bar{\rho}_k=1$ for $k<0$. Once $\mathbf{W}_k$ is obtained, the post-equalization SINR on each symbol in $\mathbf{x}_k$ can be calculated.

 Let $\mathbf{G}_{\rm DD} = \mathbf{W}_k\widetilde{\bH}_{\rm DD}^\mathrm{H}\widetilde{\bH}_{\rm DD}$. The post-equalization channel gain on element $n$ of $\mathbf{x}_k$ is given by the diagonal elements $\mathbf{G}_{\rm DD}$, i.e., $\mu_{\hat{\x}}[n] = G_{\rm DD}[n,n]$. The variance of the interference-plus-noise on element $n$ of $\mathbf{x}_k$ is calculated using the off-diagonal elements of $\mathbf{G}_{\rm DD}$ and is given by $\nu_{\hat{\x}}[n] = \sum_{m,m \not= n}|G_{\rm DD}[n,m]|^2 + C_{\rm DD}[n,n]\sigma^2$,where $\mathbf{C}_{\rm DD}=\mathbf{G}_{\rm DD}\mathbf{W}_k^\mathrm{H}.$ 

While this method provides the exact post-equalization SINR information, it is computationally expensive due to the $MN \times MN$ matrix multiplication in (\ref{eq10}). We can reduce the complexity of this computation by making the assumption that the channel matrix has a  BCCB structure. Under this assumption, $\widetilde{\bH}_{\rm DD}$ can be converted to a diagonal matrix in the TF domain via
$
     \widetilde{\bH}_{\rm TF} = (\mathbf{F}_{N}\otimes\mathbf{F}_{M})\widetilde{\bH}_{\rm DD}(\mathbf{F}_{N}\otimes\mathbf{F}_{M})^{\mathrm{H}} \label{diag_G}
$.
By using the properties of BCCB matrices \cite{Surabhi_2020}, we can also easily obtain the TF domain equivalent of $\mathbf{A}_{\rm DD}^{\rm{H}}\mathbf{A}_{\rm DD}$ as
    $\mathbf{A}_{\rm TF}^{\rm{H}}\mathbf{A}_{\rm TF} = \widetilde{\bH}_{\rm TF}^*\widetilde{\bH}_{\rm TF} + \sigma^2\mathbf{I}.$

Note that $\mathbf{W}_{1}$ is initialized as a diagonal matrix and hence, $\mathbf{W}_{k}$ retains the BCCB structure of the $\mathbf{A}_{\rm DD}^\mathrm{H}\mathbf{A}_{\rm DD}$ for $k>2$. This means that the entire recursion can be performed in the TF domain with diagonal matrices only. The recursion in (\ref{eq10}) can now be formulated in the TF domain as  
\begin{equation}
    \begin{split}
        &\mathbf{\Omega}_{k} =  \mathbf{\Omega}_{k-1} + \frac{\psi_{k-2}^2\zeta_k\Bar{\rho}_{k-3}\Bar{\phi}_{k-3}}{\zeta_{k-2}\Bar{\rho}_{k-1}\Bar{\phi}_{k-1}}\left(\mathbf{\Omega}_{k-2} -\mathbf{\Omega}_{k-3}\right) +
        \\&\left( \frac{\zeta_k\Bar{\rho}_{k-2}\Bar{\phi}_{k-2}(1+\psi_{k-1}^2)}{\zeta_{k-1}\Bar{\rho}_{k-1}\Bar{\phi}_{k-1}}\mathbf{I}_{{MN}} - \frac{\zeta_k}{\Bar{\rho}_{k-1}\Bar{\phi}_{k-1}}\mathbf{A}_{\rm TF}^{\rm{H}}\mathbf{A}_{\rm TF} \right)
        \\& \times \left(\mathbf{\Omega}_{k-1} -\mathbf{\Omega}_{k-2}\right),
    \end{split}
        \label{eq10_apx}
\end{equation}
where $\mathbf{\Omega}_{k}$ is the TF domain equivalent of $\mathbf{W}_k$ and is initialized with $\mathbf{\Omega}_1 = \frac{\zeta_1}{\Bar{\rho}_{0}\Bar{\phi}_{0}}\mathbf{I}_{MN}$ and $\mathbf{\Omega}_{k} = \mathbf{0}_{MN \times MN}$ for $k<1$. Since this recursion only involves diagonal matrices, it can be performed with a low complexity.

We can now use $\mathbf{\Omega}_{k}$ to calculate the approximate post-equalization SINR information. To this end, we calculate the TF domain equivalents of $\mathbf{G}_{\rm DD}$ and $\mathbf{C}_{\rm DD}$ as $\mathbf{G}_{\rm{TF}} = {\mathbf{\Omega}_k}\widetilde{\bH}_{\rm TF}^{\rm{H}}\widetilde{\bH}_{\rm TF}$ and $\mathbf{C}_{\rm{TF}} = \mathbf{\Omega}_k\widetilde{\bH}_{\rm TF}^{\rm{H}}\widetilde{\bH}_{\rm TF}{\mathbf{\Omega}_k^{\rm{H}}}$, respectively. 
The reverse process in obtaining $\widetilde{\bH}_{\rm TF}$ from $\widetilde{\bH}_{\rm DD}$ can then be used to calculate approximations of the DD domain matrices $\mathbf{G}_{\rm DD}$ and $\mathbf{C}_{\rm DD}$, i.e.,
    $
    \widetilde{\bm{G}}_{\rm DD} = (\mathbf{F}_{N}\otimes\mathbf{F}_{M})^{\mathrm{H}}\mathbf{G}_{\rm{TF}}(\mathbf{F}_{N}\otimes\mathbf{F}_{M})$
    and
    $\widetilde{\bm{C}}_{\rm DD} = (\mathbf{F}_{N}\otimes\mathbf{F}_{M})^{\mathrm{H}}\mathbf{C}_{\rm{TF}}(\mathbf{F}_{N}\otimes\mathbf{F}_{M}).$
Since $\widetilde{\bm{G}}_{\rm DD}$ and $\widetilde{\bm{C}}_{\rm DD}$ are BCCB matrices, their respective rows are simply shifted versions of each other. Therefore, under this approximation, each symbol experiences the same SINR, the post-equalization channel gain is given by 
\vspace{-1ex}
\be
\tilde{\mu}_{\hat{\x}}[n] = \widetilde{G}_{\rm DD}[1,1], \label{mean_apx}
\vspace{-1ex}
\ee
and the variance of the interference-plus-noise is given by
\vspace{-0.5ex}
\be
\tilde{\nu}_{\hat{\x}}[n] = \sum_{m=2}^{MN-1}|\widetilde{G}_{\rm DD}[1,m]|^2 + \widetilde{C}_{\rm DD}[1,1]\sigma^2. \label{var_apx}
\vspace{-1ex}
\ee
\subsection{LLR Computation}
\vspace{-1ex}
After performing symbol-wise de-interleaving upon the estimated symbol vector output from the mLSQR algorithm, the soft symbol demapper computes the extrinsic information of the coded bits in terms of log-likelihood ratio.  The extrinsic LLR, $L_{\rm ext}[c_{n,k}], \, 0\leq n \leq MN-1, \, 0\leq k \leq Q-1$,  is the information of the coded bits $c_{n,k}$ contained in $\bm Y$ and the \textit{a-priori} information of $c_{n',k}$, $\forall n' \neq n $.  This extrinsic information is then fed to the decoder  which computes the \textit{a-posteriori} information. During the $i^{\text{th}}$ iteration of TTE-SIC, the mLSQR algorithm computes the estimate $\hat{\x}^i \in \mathbb{C}^{MN\times 1}$,  the mean $\tilde{\bmu}^i_{\hat{\x} }\in \mathbb{C}^{MN\times 1}$ and the variance $\tilde{\bnu}^i_{\hat{\x} }\in \mathbb{C}^{MN\times 1}$.  Using the output from mLSQR, we can compute the extrinsic information of the coded bits as  
\vspace{-0.5ex}
\begin{align}\nonumber
&L^i_{\rm ext}\left[ c_{n,k}\vert \hat{x}^i [n] \right]=\\
&{\rm ln}\left(\frac{\underset{\substack{\forall \balpha_l:\\ \alpha_{l,k}=0}}{\sum}P(\hat{x}^i [n]\vert \bc_n=\balpha_l )\underset{\substack{\forall k': k'\neq k }}{\prod}P(c_{nm'}=\alpha_{lm'})}{\underset{\substack{\forall \balpha_l:\\ \alpha_{l,k}=1}}{\sum}P(\hat{x}^i [n]\vert \bc_n=\balpha_l )\underset{\substack{\forall k': k'\neq k }}{\prod}P(c_{nm'}=\alpha_{lm'})}\right),\label{eqn:l_ext}
\vspace{-0.5ex}
\end{align}
where $\bc_n=[c_{n0},c_{n1},\hdots, c_{nm},\hdots, c_{nQ-1}]^{\rm T}$, $\balpha_l=[\alpha_{l0},\alpha_{l1},\hdots, \alpha_{lm},\hdots, \alpha_{lQ-1}]^{\rm T}$ with $c_{nm}, \alpha_{lm} \in \{0,1\},\, 0\leq l\leq 2^Q-1$ \cite{Tuchler_2002}.
In order to simplify the LLR derivation, we assume 
$P(\hat{x} ^i[n] \vert \bc_n=\balpha_l)\sim \mathcal{CN}(q_l\tilde{\mu}^i_{\hat{x}}[n],\tilde{\nu}^i_{\hat{x} }[n])$, where $q_l$ is the modulated symbol in symbol constellation corresponding to the binary vector $\balpha_l$. This assumption is widely used in turbo equalization methods \cite{Xiaodong_1999}. Hence, (\ref{eqn:l_ext}) is simplified as 
\begin{align}\nonumber
&L^i_{\rm ext}\left[c_{n,k}\vert \hat{x} [n] \right]= \\
&{\rm ln}\!\left(\frac{\underset{q\in\mathcal{Z}_k^0}{\sum}{\rm exp}\!\left(-\frac{\vert \hat{x}^i[n]-q\tilde{\mu}^i_{\hat{x}}[n]\vert^2}{\tilde{\nu}^i_{\hat{\x}}[n]}\right)\!\!+\!\!\!\underset{\substack{\forall k'\\ k'\neq k}}{\sum}\!\frac{1-2\alpha_{q,k'}}{2}L[c_{n,k'}]}{\underset{q\in\mathcal{Z}_k^1}{\sum}{\rm exp}\!\left(-\frac{\vert \hat{x}^i [n]-q\tilde{\mu}^i_{\hat{\x}}[n]\vert^2}{\tilde{\nu}^i_{\hat{\x}}[n]}\right)\!\!+\!\!\!\underset{\substack{\forall k'\\ k'\neq k}}{\sum}\!\frac{1-2\alpha_{q,k'}}{2}L[c_{n,k'}]} \right)\!,\label{eqn:l_ext2}
\end{align}
where $\mathcal{Z}_k^0$ and $\mathcal{Z}_k^1$ are the sets of modulated symbols corresponding to input signal vector with $0$ and $1$ in the $k^{\text{th}}$ position, respectively. In (\ref{eqn:l_ext2}), $\alpha_{q,k'}$ is the bit $k'$ in the input vector which is mapped to $q$. 
Furthermore, using max-log approximation, \cite {Studer_2011_ASIC}, and omitting the constant terms, $L^i_{\rm ext}\left[c_{n,k}\vert \hat{x} [n] \right]$ can be computed as, 
\begin{align}\label{eqn:l_ext3}
L^i_{\rm ext}&\left[c_{n,j}\vert \hat{x}^i [n] \right]\!\!=\!\!\!\!\!\!
&\gamma_n\!\!\left(\underset{q\in \mathcal{Z}_k^0}{\rm min}\vert \hat{x}^i [n]-q\vert^2 \!\!-\!\!\underset{q\in \mathcal{Z}_k^1}{\rm min} \vert \hat{x}^i [n]-q\vert^2\right)
\end{align}
where the SINR at the output of equalizer is defined as $\gamma_n=\frac{(\tilde{\mu}^{i}_{\hat{\x}}[n])^2}{\tilde{\nu}^i_{\hat{\x}}[n]}$.
This extrinsic information is fed to the decoder, which computes the \textit{a-posteriori} information of the coded bits, $\lambda^i[c_{n,k}\vert \hat{x}^i [n] ]$.  The \textit{a-priori} information of each code bit can then be calculated as 
\begin{align}\label{eqn:l_apr}
\lambda^i[c_{n,k}]=\lambda^i\left[c_{n,k}\vert \hat{x}^i [n] \right]-L^i_{\rm ext}\left[c_{n,k}\vert \hat{x} ^i[n] \right].
\end{align}
The \textit{a-priori} information of all encoded bits is then collected in a vector ${\blambda}^i \in \mathbb{R}^{MNr}$ to perform interference cancellation.

\begin{algorithm}[t]
\caption{Proposed TTE-SIC Technique}\label{turbo}
\begin{algorithmic}[1]
\State \textbf{Input}: $\bm{y}_{\rm DD}$, $\widetilde{\bH}_{\rm DD}$ and $\bDelta_{\rm DD}$
\State \textbf{Initialize}: $\bmu^{0'}_{\hat{\x}}=\bm 0$ and $\bm{y}_{\rm DD}^0=\bm{y}_{\rm DD}$
\For{$i=1:I_{\rm SIC}^{\rm Prop.}$}
\State Calculate $\bm{y}_{\rm DD}^{i+1}$ using (\ref{eqn:ici_tf})
\State Obtain  $\hat{\x}^i$, $\tilde{\bmu}_{\hat{\x} }^i$ and $\tilde{\bnu}_{\hat{\x}^i }$ using  Algorithm \ref{LSQR} 
\State Calculate $L^i_{\rm ext}\left[c_{n,k}\vert \hat{x}^i [n] \right]$ as in (\ref{eqn:l_ext3})
\State Calculate $\lambda^i[c_{n,k}]$ using (\ref{eqn:l_apr})
\State Obtain the soft output, $\mu_{\hat{\x} }^{i'}[n]$ using (\ref{eqn:soft_op})
\EndFor
\State \textbf{Output}: Obtain $\hat{\bm{b}}$ from decoder output
\end{algorithmic}
\end{algorithm}
\subsection{Interference Cancellation}
Using the \textit{a-priori} information of the codewords, the soft symbol estimates can be obtained. After performing bitwise deinterleaving on ${\blambda}^i $, the soft symbol mapper calculates the new refined soft symbol estimates as  
\begin{align}\label{eqn:soft_op}
\mu_{\hat{\x} }^{i'}[n]=\sum_{q\in\mathcal{Z}}qP(\hat{x}^i [n]=q),
\end{align}
where  $P(\hat{x}^i [n]=q)\!=\!\prod_{j=0}^{Q-1}\frac{1}{2}\left(1\!+\!(\tilde{\alpha}_{q,j}){\rm tanh}(\frac{\lambda^i[c_{n,j}]}{2}) \right)$, with $\tilde{\alpha}_{q,j}=2\alpha_{q,j}\!-\!1$.
We then use the soft symbol estimates from (\ref{eqn:soft_op}) to remove the interference caused by the \textit{insignificant} DD domain channel coefficients which remains in the received signal. The received signal at iteration $i+1$ after interference cancellation is expressed as 
\begin{align}\label{eqn:ici}
\mathbf{y}^{i+1}_{\rm DD}=\mathbf{y}^i_{\rm DD}-\bDelta_{\rm DD}\bmu^{i'}_{\hat{\x}},
\end{align}
where $\bmu^{i'}_{\hat{\x}}=[\mu^{i'}_{\hat{\x}}[0], \ldots, \mu^{i'}_{\hat{\x}}[MN-1]]^{\rm T}$. 
The computational complexity of this interference cancellation procedure can be reduced by implementing it in the TF domain. Since $\bDelta_{\rm DD}$ has a BCCB structure, each of the $N-B'$ circulant $M \times M$ sub-matrices can be diagonalized using an $M-$ point DFT, where $B'=2B+1$. Therefore, the interference cancellation in (\ref{eqn:ici}) can be performed in TF domain as
\begin{align}\label{eqn:ici_tf}
\mathbf{y}^{i+1}_{\rm TF}=\mathbf{y}_{\rm TF}^i-{\bDelta}_{\rm TF}\tilde{\bmu}^{i'}_{\hat{\x}}
\end{align}
where ${\bm{y}}^i_{\rm TF}$, $\bDelta_{\rm TF}$ and $\tilde{\bmu}^{i'}_{\hat{\x}}$ are obtained from ${\bm{y}}^i_{\rm DD}$, $\bDelta_{\rm DD}$ and $\bmu^{i'}_{\hat{\x}}$ respectively by multiplying  them with $(\bF_{N}\otimes\bF_{ M})^{}$ from left and $(\bF_{ N}\otimes \bF_M)^{\rm H}$ from right. 
Since $\bDelta_{\rm TF}$  can be pre-computed, the interference cancellation at iteration $i$ can be performed with $MN(N-B')$ element-wise complex multiplications and ${\bm{y}}^{i+1}_{\rm DD}=(\bF_{N}\otimes\bF_{ M})^{\rm H}{\bm{y}}_{\rm TF}^{i+1}(\bF_{N}\otimes\bF_{ M})$. Finally, our proposed technique is summarized in Algorithm~\ref{turbo}. 

\subsection{Computational Complexity}\label{sec:complexity}
%

\begin{figure}[t]
    \centering
    \vspace{-0.4cm}
    \includegraphics[width=0.7\columnwidth]{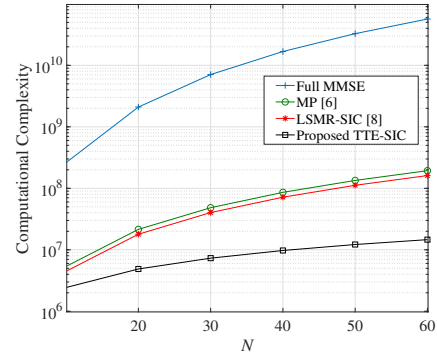}
    \vspace{-0.1cm}
    \caption{Computational complexity comparison of our proposed TTE-SIC technique with the existing methods.}
    \label{fig:complexity}
\end{figure}

%
In this section, we analyze and compare the complexity of the proposed TTE-SIC with the existing methods in literature in terms of the number of CMs. We consider the full MMSE equalizer without channel truncation followed by a decoder as a benchmark which has the computational load of $\mathcal{O}(M^3N^3)$. The computational complexity order for mLSQR  is $\mathcal{O}(MNP'I_{\rm LSQR})$, where $P'$ is the number of non-zero elements in a column of $\bH_{\rm DD}$ and $I_{\rm LSQR}$ is the number of LSQR iterations\cite{ Hrycak_LSQR_GMRES}. In a scattering rich environment with fractional Doppler shifts, $P'=NL$, and hence, the complexity order will be $\mathcal{O}(MN^2LI_{\rm LSQR})$. Using the truncated channel, LSQR complexity in step 5 of TTE-SIC in Algorithm~\ref{turbo} is $\mathcal{O}(MNBLI_{\rm LSQR})$.
The soft symbol mapping and demapping in steps 6, 7 and 8 of Algorithm \ref{turbo} require a total of $2(Q+1)MN$ CMs. 
Furthermore, by performing the interference cancellation in TF domain, step 4 requires $(N-B')\mathcal{O}(M\log_2(M))+MN(N-B')+\mathcal{O}(MN\log_2(MN))$ CMs. The MP equalizer in \cite{Raviteja_2018} and the LSMR with SIC receiver in \cite{Qu_lsmr} require $\mathcal{O}(MN^2LQI_{\rm MP})$ and $\mathcal{O}(MN^2LI_{\rm LSMR}I_{\rm SIC}^{\rm LSMR.})$ CMs, respectively. The MP detector requires $I_{\rm MP}$ iterations and  LSMR with SIC receiver requires  $I_{\rm LSMR}$ iterations for LSMR and $I_{\rm SIC}^{\rm LSMR}$ iterations for SIC. The computational complexities of different techniques are summarised in Table \ref{tab:complexity}. It should be noted that, the equalizers in \cite{Raviteja_2018} and \cite{Qu_lsmr} do not consider the soft information available at the output of the decoder. Therefore, for a fair comparison, the computational complexity associated with the decoders are not included. Fig.~\ref{fig:complexity} shows the computational load of the proposed TTE-SIC receiver compared to different methods from the literature as a function of $N$. For this comparison, we consider $L=7$, $Q=4$, $I_{\rm MP}=30$, $I_{\rm LSMR}=I_{\rm LSQR}=20$, $I_{\rm SIC}^{\rm LSMR}=5$ \cite{Qu_lsmr}. For the proposed TTE-SIC technique, we used $B=2$, and $I_{\rm SIC}^{\rm Prop.}=3$ for SIC. As seen in Fig.~\ref{fig:complexity}, our proposed technique is around $10^4$ and $10$ times simpler than full MMSE and LSMR-SIC equalizers, respectively.

\begin{table}
  \centering
    \caption{Computational Complexity Comparison}
    \label{tab:complexity}
  \resizebox{1\columnwidth}{!}
{\begin{tabular}{|c|c|}
\hline\hline
Equalization Methods  & Complexity Order  \\ \hline\hline
Full MMSE & $\mathcal{O}(M^3N^3)$\\ \hline
MP \cite{Raviteja_2018}& $\mathcal{O}(MN^2LQI_{\rm MP})$  \\ \hline
LSMR with SIC \cite{Qu_lsmr}& $\mathcal{O}(MN^2LI_{\rm lSMR}I_{\rm SIC}^{\rm LSMR})$\\ \hline
Proposed TTE-SIC&\shortstack[]{ $\mathcal{O}(MNI_{\rm SIC}^{\rm Prop.}(B'LI_{\rm LSQR}+2Q+N-B'+4-\frac{B'}{N}\log_2M+\log_2N))$} \\
 \hline\hline
    \end{tabular}}
    \vspace{-0.4 cm}
\end{table}

\vspace{-0.2cm}
\section{Simulation Results}\label{sec:simultion}
\begin{figure*}[t]
\centering
\begin{subfigure}{1\columnwidth}
	\centering
	\includegraphics[width=0.85\columnwidth]{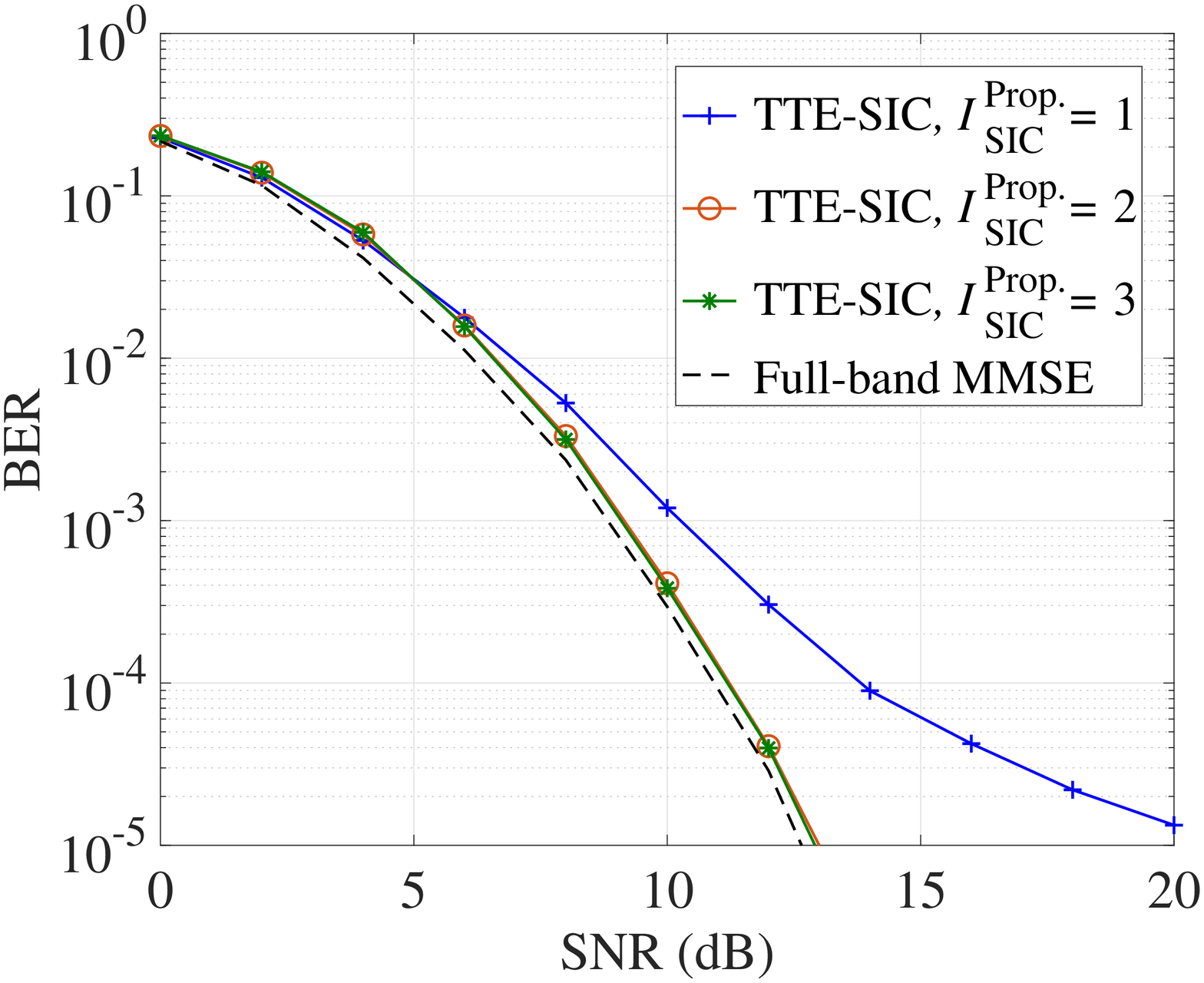}
	\caption{$B=2$}
	\label{fig:vel_500_1}
\end{subfigure}
\begin{subfigure}{1\columnwidth}
	\centering
	\includegraphics[width=0.85\columnwidth]{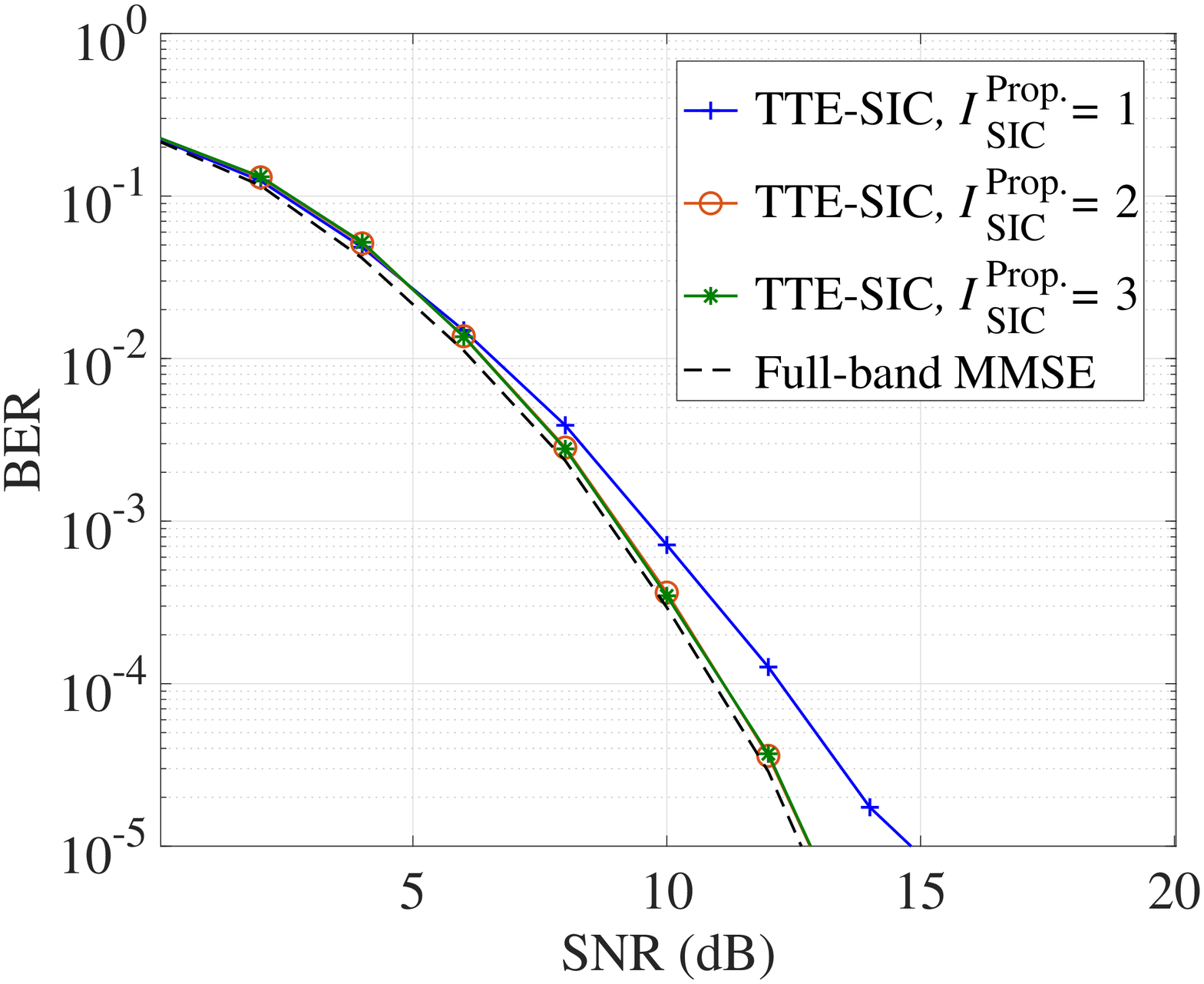}
	\caption{$B=3$}
	\label{fig:vel_500_2}
\end{subfigure}
\caption{BER performance of the proposed TTE-SIC technique at a relative velocity of $500$~km/h.}
\label{fig:vel_500}
\vspace{-0.5 cm}
\end{figure*}
In this section, we evaluate the performance of our proposed TTE-SIC technique via simulations. We consider an OTFS system with $M=64$ and $N=16$. At the transmitter, the information bits are encoded using a $1/2$ rate convolutional encoder with $(2,1,3)$ configuration and are mapped on to $4-$QAM constellation. In our simulations, we use the extended vehicular A (EVA) channel model, \cite{3gppr12}, with a carrier frequency of $f_c = 5.9$~GHz, a sampling period of $T_s = 370.3$~ns and a relative velocity of $500$~km/h between the transmit and receive antennas which corresponds to $f_{\rm Dmax}=2.73$~kHz. We set the MMSE equalizer with the full channel, i.e., no truncation, as a benchmark in our evaluations.


Fig.~\ref{fig:vel_500} shows the BER performance of our proposed TTE-SIC method versus SNR for different values of SIC iterations and the parameter $B$. Fig.~\ref{fig:vel_500_1} shows the BER performance for the channel truncation bandwidth given by the criterion described in Section~\ref{sec_truncation}, which corresponds to $B= \lceil f_{\rm Dmax}MNT_{\rm s}\rceil = 2$. It can be seen that, only 2 iterations are sufficient for our proposed technique to provide about the same performance as that of our benchmark. In Fig.~\ref{fig:vel_500_2}, we repeat our evaluations for $B=3$ to asses the validity of our channel truncation criterion. This analysis reveals that increasing $B$ brings only marginal performance gain when $I^{\rm Prop.}_{\rm SIC}=1$ and negligible gain when $I^{\rm Prop.}_{\rm SIC}\geq 2$. This confirms that our criterion defines sufficient level of channel truncation.
Based on the results in Fig.~\ref{fig:vel_500} and Section~\ref{sec:complexity}, our proposed TTE-SIC technique achieves a similar BER performance to the benchmark with several orders of magnitude lower computational load.

\section{Conclusion}\label{sec:conclusion}
\vspace{0 cm}
In this paper, we have proposed a novel low-complexity TTE-SIC receiver for coded OTFS systems in high-mobility scenarios. In our proposed technique, we exploit the block-banded structure of the DD domain channel to truncate the channel and reduce the complexity of equalization.  Our proposed TTE-SIC technique deploys the low-complexity mLSQR algorithm for both channel equalization and to obtain the post-equalization SINR necessary for computing the LLRs for soft-output detection. To compensate for the performance loss due to the channel truncation, we take a low-complexity interference cancellation approach using the decoder output and insignificant Doppler coefficients left over from the truncation. Our simulation results demonstrate that, after only 2 iterations, the proposed TTE-SIC technique offers about the same performance as the full-channel MMSE benchmark with orders of magnitude lower computational load. 


\bibliographystyle{IEEEtran}
\bibliography{IEEEabrv,references}
\end{document}